\RequirePackage[2020-02-02]{latexrelease}
\documentclass[aps,prc,twocolumn,showpacs,showkeys,nofootinbib,superscriptaddress]{revtex4-2}

\usepackage{bm}
\usepackage{longtable}
\usepackage{graphicx}
\usepackage{amsmath}
\usepackage{amsfonts}
\usepackage{amssymb}
\usepackage{natbib} 
\usepackage{setspace}
\usepackage{xspace}
\usepackage{cancel}
\usepackage{physics}
\usepackage{tensor,braket,color,bbm,slashed}

\usepackage{newtxtext,newtxmath}
\usepackage{mathrsfs}

\usepackage[normalem]{ulem}
      
\renewcommand{\sout}{\bgroup \color{red} \ULdepth=-.5ex \ULset}

\begin{document}
\title{
  \begin{flushright}
    \rightline{PKNU-NuHaTh-2021-08} 
  \end{flushright}
  Implications of PREX--2 data on the electron--neutrino opacity in dense matter}

\author{Parada T.P. Hutauruk}
\email{phutauruk@gmail.com}
\affiliation{Department of Physics, Pukyong National University (PKNU), Busan 48513, Korea}

\date{\today}

\begin{abstract}
Motivated by the recent measurement of the neutron distribution radius of ${}^{208}$Pb from the PREX--2 data, I study the effects of the new G3(M) parameter set constrained by PREX--2 data on the electron--neutrino scattering in dense matter using the extended relativistic mean field (E--RMF) model. I employ the G3(M) parameter set to describe the nuclear matter. The obtained equation of state for the G3(M) parameter set has an excellent agreement with experimental data and the chiral effective field theory calculation with N$^3$LO 3N forces. I analyze both differential cross section of electron--neutrino and electron--neutrino mean free path to observe their sensitivity to the G3(M) parameter set. One finds that the differential cross sections of electron--neutrino for different baryon densities have higher values compared with that obtained for the TM1e and FSU Garnet parameter sets. The higher cross section decreases the electron--neutrino mean free path.     
\end{abstract}



\maketitle

\section{Introduction} \label{intro}

Neutrinos play very important role in the evolution processes of the neutron star (NS) and supernova, which correspond with the cooling rate of neutron star, that controlled by neutrinos emission~\cite{Yakovlev:2000jp,Burrows:1987zz}. The neutrinos emission and scattering are sensitive to the equation of state (EOS) of nuclear matter~\cite{Hutauruk:2020mhl,Hutauruk:2019ptu,Hutauruk:2018cgu,Reddy:1997yr,Niembro:2001hd} and the nucleon effective mass $M_N^{*}$~\cite{Niembro:2001hd}.

Recently, a measurement of parity violating asymmetry $A_{PV}$ at transferred momentum $\textbf{q} = 0.3978/$fm in ${}^{208}$Pb by Lead Radius Experiment (PREX--2) reported an accurate determination of the neutron skin thickness of ${}^{208}$Pb with a precision of approximately equal 1\%~\cite{Adhikhari21}. Combining analysis of PREX--2~\cite{Adhikhari21} with the PREX previous measurement~\cite{Abrahamyan2012} gives
\begin{eqnarray}
  \Delta R_{\textrm{skin}} &=&  R_n - R_p = (0.283 \pm 0.071)~\textrm{fm},
\end{eqnarray}
where $R_n$ and $R_p$ are respectively the root--mean--squared radii for the neutron and proton density distributions. The accurate measurement of neutron skin thickness ($\Delta R_{np} = R_n -R_p$) is very important and useful quantity to constrain the EOS for finite nuclei and nuclear matter. The reliable EOS can be obtained by refitting the appropriate parameters within the theoretical models to reproduce the properties of finite nuclei. The better constraint on the EOS in particular at high baryon density will lead us to gain a deeper understanding of the properties of neutron stars such as the size, mass, and particle composition of $\beta$--stable matter. Interestingly, it may also affect the electron--neutrino scattering with matter inside NS. Henceforth electron--neutrino is simply referred as ``neutrino''.

In recent works, several attempts have been made to investigate the implications of PREX--2 data on the EOS~\cite{PPBP21}, symmetry energy~\cite{RFHP21}, and electric dipole polarizability~\cite{Piekarewicz21}. In addition, the implications of new EOS constrained by PREX--2 data have been applied to study the properties of neutron star~\cite{PPBP21,RFHP21}. A very recent work of Ref.~\cite{PPBP21} has developed new parameter set of the E--RMF model through the fine--tuning parameters ($\Lambda_\omega$, which relates with $\eta_{2\rho}$ in the E--RMF model of the present work, and $g_\rho$ parameters) to fit the $R_n$ of PREX--2 data~\cite{Adhikhari21}. The new parameter set is labeled as the G3(M) parameter set. Thus, it was used to validate the constraint from GW170817 binary NS merger to understand the properties of neutron star. The G3(M) parameter set relatively gives a better prediction for finite and nuclear matter. Further detailed explanations of this G3(M) parameter set can be found in Ref.~\cite{PPBP21}. So far, in the literature, this G3(M) parameter set is not yet applied to the neutrino scattering with dense matter. Therefore, it will be very interesting and challenging to investigate how sensitive the EOS and neutrino scattering observable to the new G3(M) parameter set constrained by an accurate measurement of the PREX--2 data~\cite{Adhikhari21}.

In present work, I perform the extended relativistic mean--field model with modified G3(M) parameter set~\cite{PPBP21} constrained by the PREX--2 data~\cite{Adhikhari21}. The E--RMF model has been widely used to study the finite nuclei and infinite nuclear matter~\cite{Kumara:2017bti,Agrawal:2012rx,Furnstahl:1995zb}. The predictions of this model are relatively good for describing the bulk properties of finite nuclei at saturation density and properties of neutron star. In this work, I calculate the EOS, particle fractions of the constituents of $\beta$--stable matter, which consist of electrons, neutrons, protons and muons, differential cross section (DCRS) of neutrino and neutrino mean free path (NMFP). I then observe the sensitivity of these quantities to the G3(M) parameter set.

One finds that the binding energy per nucleon $E_B/A$ for pure neutron matter (PNM) with the G3(M) parameter set is softer than that obtained for the TM1e~\cite{Bao:2014lqa} and FSU Garnet~\cite{CP14} parameter sets at low baryon density. However, the $E_{B}/A$ for the G3(M) parameter set is rather stiffer than that obtained for the FSU Garnet parameter set and the same as that obtained for the TM1e parameter set. The PNM pressure for the G3(M) parameter set fits well with the asy--soft of the flow data~\cite{Danielewicz:2002pu} at intermediate $\rho_B/\rho_0$ with $\rho_B$ and $\rho_0$ are respectively the baryon and saturation densities. It is in good agreement with the asy--stiff of the flow data~\cite{Danielewicz:2002pu} at higher $\rho_B/\rho_0$. The sound velocities $v_s (c)$ for the G3(M), TM1e and FSU Garnet parameter sets predict the same sound velocity at around saturation density ($\rho_B / \rho_0 \simeq $ 1) and at $\rho_B / \rho_0 \simeq$ 3.7.

Total differential cross sections of neutrino for the GM(3) parameter set are found to be higher than that obtained for the TM1e and FSU Garnet parameter sets for different baryon densities. As consequences, the NMFP for the G3(M) parameter set is lower than obtained for the TM1e and FSU Garnet parameter sets. The higher DCRS or lower NMFP for the G3(M) parameter set is expected due to the nucleon effective mass $M_N^*$ is higher than that obtained for FSU Garnet and TM1e parameter sets.

This paper is organized as follows. In Sec.~\ref{sec:eos} I briefly introduce the effective Lagrangian for nuclear matter within the E--RMF model with the G3(M) parameter set. I then calculate the EOS like the binding energy, pressure, and sound velocity for pure neutron matter. In Sec.~\ref{sec:interaction} I present the expression for both differential cross section of neutrino and neutrino mean free path and I then observe their sensitivity to the G3(M) parameter set as well as the nucleon effective mass. In Sec.~\ref{results} the results are presented and their implications are discussed. Sec.~\ref{summary} is devoted for a summary.
%

\section{E-RMF model}
\label{sec:eos}
%
%

Here I briefly introduce the EOS of dense matter that used to describe the constituents of matter. I employ the E--RMF model with the modified G3(M) parameter set, as mentioned above already. The effective Lagrangian for the E--RMF model is given by~\cite{Furnstahl:1995zb,Furnstahl:1996wv}
\begin{eqnarray}
  \label{eq:eqxenont1}
  \mathscr{L}_{ERMF} &=& \mathscr{L}_{NM} + \mathscr{L}_{\sigma}
  + \mathscr{L}_{\omega} + \mathscr{L}_{\rho}
  + \mathscr{L}_{\delta}
  + \mathscr{L}_{\sigma \omega \rho},  
\end{eqnarray}
where the interaction Lagrangian of the nucleons and mesons is defined by
\begin{eqnarray}
  \label{eq:eqxenont2}
  \mathscr{L}_{NM} &=& \sum_{j=n,p} \bar{\psi}_j \left[ i \gamma^\mu \partial_\mu
    - (M_N - g_\sigma \sigma - g_\delta \bm{\tau}_j \cdot \bm{\delta}) \right. \nonumber \\ && \left. \mbox{}
    - \left( g_\omega \gamma^\mu \omega_\mu
    + \frac{1}{2} g_\rho \gamma^\mu \bm{\tau}_j \cdot \bm{\rho}_\mu \right) \right] \psi_j ,
\end{eqnarray}
where the sum stands for the neutrons and protons, $M_N$ is the nucleon mass, and $\bm{\tau}_j$ are the isospin matrices. The $g_\sigma$, $g_\omega$, $g_\rho$, and $g_\delta$ are respectively the coupling constants for the $\sigma$, $\omega$, $\rho$, and $\delta$ mesons. The self--interactions Lagrangian for the $\sigma$, $\omega$, $\rho$, and $\delta$ mesons are expressed by
\begin{eqnarray}
  \label{eq:eqxenon3}
  \mathscr{L}_{\sigma} &=& \frac{1}{2} \left( \partial_\mu \sigma \partial^\mu \sigma -m_\sigma^2 \sigma^2 \right)
  - \frac{\kappa_3}{6M_N} g_\sigma m_\sigma^2 \sigma^3 \nonumber \\
  &-& \frac{\kappa_4}{24 M_N^2} g_\sigma^2 m_\sigma^2 \sigma^4 , \\
  \mathscr{L}_{\omega} &=& - \frac{1}{4} \omega_{\mu \nu} \omega^{\mu \nu} + \frac{1}{2} m_\omega^2 \omega_\mu \omega^\mu
  + \frac{1}{24} \xi_0 g_\omega^2 (\omega_\mu \omega^\mu)^2 , \\
  \mathscr{L}_\rho &=& -\frac{1}{4} \bm{\rho}_{\mu \nu} \cdot \bm{\rho}^{\mu \nu} + \frac{1}{2} m_\rho^2 \bm{\rho}_\mu \cdot \bm{\rho}^\mu, \\
  \mathscr{L}_\delta &=& \frac{1}{2} \partial_\mu \bm{\delta} \cdot \partial^\mu \bm{\delta} - \frac{1}{2} m_\delta^2 \bm{\delta}^2.
\end{eqnarray}
The $m_\sigma$, $m_\omega$, $m_\rho$ and $m_\delta$ are the meson masses. The $\omega^{\mu \nu}$ and $\rho^{\mu \nu}$ are the field tensors for the $\omega$ and $\rho$ mesons, which are respectively defined as $\omega^{\mu \nu} = \partial^\mu \omega^\nu - \partial^\nu \omega^\mu$, and $\bm{\rho}^{\mu \nu} = \partial^\mu \bm{\rho}^\nu - \partial^\nu \bm{\rho}^\mu - g_\rho (\bm{\rho}^\mu \times \bm{\rho}^\nu)$.

The non--linear cross interaction Lagrangian of $\sigma$, $\omega$ and $\rho$ mesons is given by
\begin{eqnarray}
  \label{eq:eqxenont4}
  \mathscr{L}_{\sigma \omega \rho} &=&
  \frac{\eta_1}{2M_N} g_\sigma m_\omega^2 \sigma \omega_\mu \omega^\mu
  + \frac{\eta_2}{4M_N^2} g_\sigma^2 m_\omega^2 \sigma^2 \omega_\mu \omega^\mu \nonumber \\
  &+& \frac{\eta_\rho}{2M_N} g_\sigma m_\rho^2 \sigma \bm{\rho}_\mu \cdot \bm{\rho}^\mu
  + \frac{\eta_{1\rho}}{4 M_N^2} g_\sigma^2 m_\rho^2 \sigma^2 \bm{\rho}_\mu \cdot \bm{\rho}^\mu \nonumber \\
  &+& \frac{\eta_{2 \rho}}{4 M_N^2} g_\omega^2 m_\rho^2 \omega_\mu \omega^\mu \bm{\rho}_\mu \cdot \bm{\rho}^\mu ,
\end{eqnarray}
where $\kappa_3$, $\kappa_4$, $\xi_0$, $\eta_1$, $\eta_2$, $\eta_\rho$, $\eta_{1\rho}$ and $\eta_{2 \rho}$ are the coupling constants. The complete values of the coupling constants in the Lagrangian in Eqs.~(\ref{eq:eqxenont2})--(\ref{eq:eqxenont4}) are summarized in Table~\ref{tab:model1}.

The Lagrangian density for the electron and muon is given by
\begin{eqnarray}
  \mathscr{L}_l &=& \sum_{l =e,\mu} \bar{\psi}_l (i\gamma_\mu \partial^\mu - m_l) \psi_l,
\end{eqnarray}
with $m_l$ is the lepton mass. Once the Lagrangian is given, the particle composition of the dense matter with $n$, $p$, $e$ and $\mu$ can be obtained through the constraint of $\beta$--equilibrium that states the relation of the chemical potential as
\begin{eqnarray}
  \mu_n - \mu_p &=& \mu_e, \hspace{1.0cm} \mu_e = \mu_\mu,
\end{eqnarray}
and the charge neutrality,
\begin{eqnarray}
  \rho_p &=& \rho_e + \rho_\mu,
\end{eqnarray}
where the leptons are treated as relativistic ideal Fermi gases. The $\mu_{p,n} = \frac{\partial \mathscr{E}}{\rho_{p,n}}$ are the chemical potentials of proton and neutron, respectively and the $\rho_p$, $\rho_e$ and $\rho_\mu$ are respectively proton, electron and muon densities. The chemical potentials for lepton are given by $\mu_{l=e,\mu} = \sqrt{k_{F_{l=e,\mu}}^2 + m_{l=e,\mu}^2}$. The total baryon density is defined by $\rho_B = \rho_p + \rho_n$ with $\rho_n$ is neutron density.

\begin{figure}[t]
  \centering\includegraphics[width=1\columnwidth]{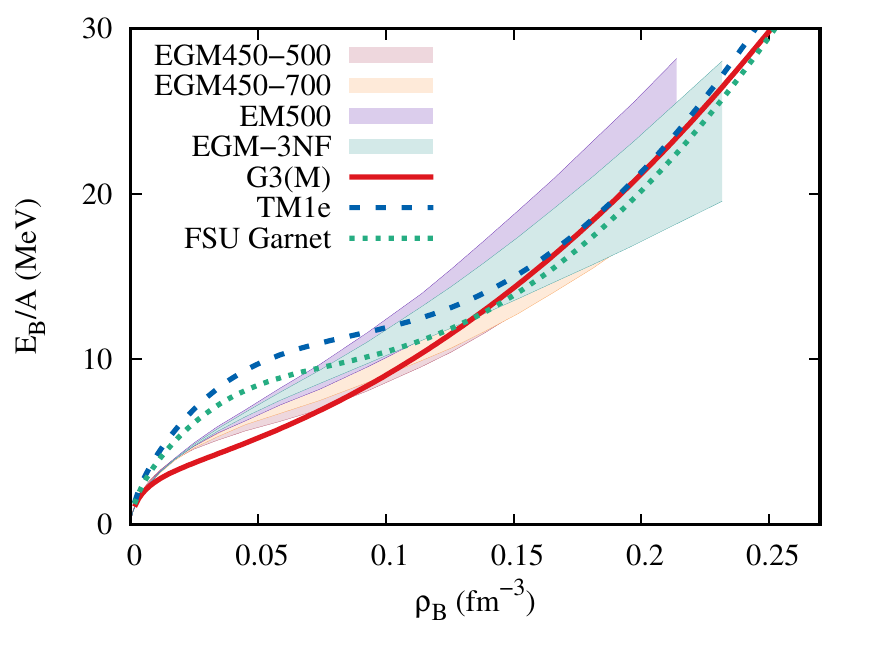}
  \caption{\label{fig1} (Color online) Binding energy for pure neutron matter as a function of $\rho_B$ calculated in the E--RMF model for the G3(M)~\cite{PPBP21} (solid line), TM1e~\cite{Bao:2014lqa} (dashed line) and FSU Garnet~\cite{CP14} (dotted line) parameter sets. The chiral effective field theory calculation with N$^3$LO 3N forces is taken from Ref.~\cite{Tews:2012fj} (green shaded area).}
\end{figure}

\begin{figure}[t]
  \centering\includegraphics[width=1\columnwidth]{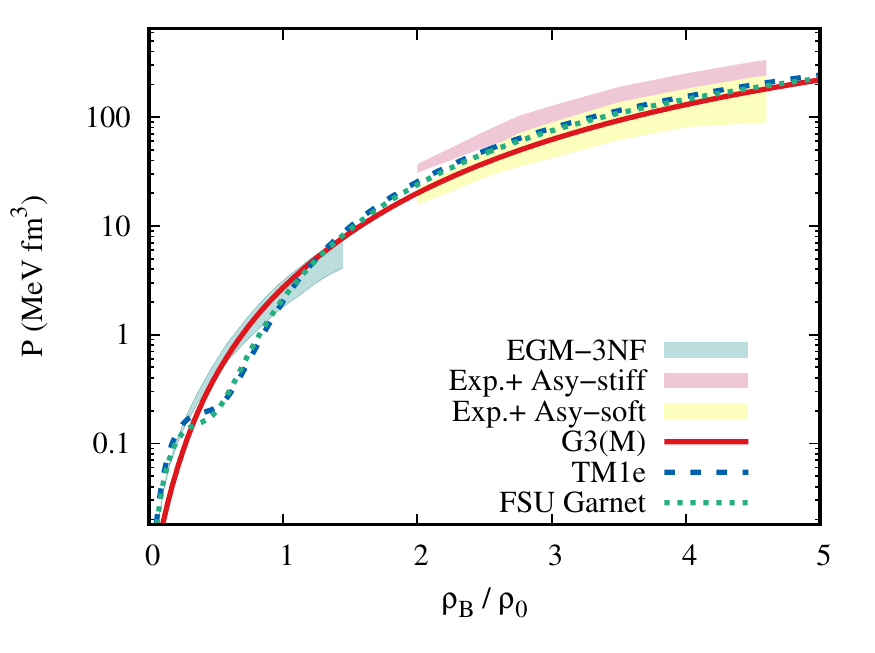}
  \caption{\label{fig2} (Color online) Pressure for pure neutron matter as a function of $\rho_B /\rho_0$ calculated in the E--RMF model using the same parameter sets as used in Fig.~\ref{fig1}. The line representations are the same as in Fig.~\ref{fig1}. The chiral effective field theory calculation with N$^3$LO 3N forces is taken from Ref.~\cite{Tews:2012fj} (green shaded area). The yellow and pink shaded area are experimental data, which is taken from Ref.~\cite{Danielewicz:2002pu}.}
\end{figure}

\begin{figure}[t]
  \centering\includegraphics[width=1\columnwidth]{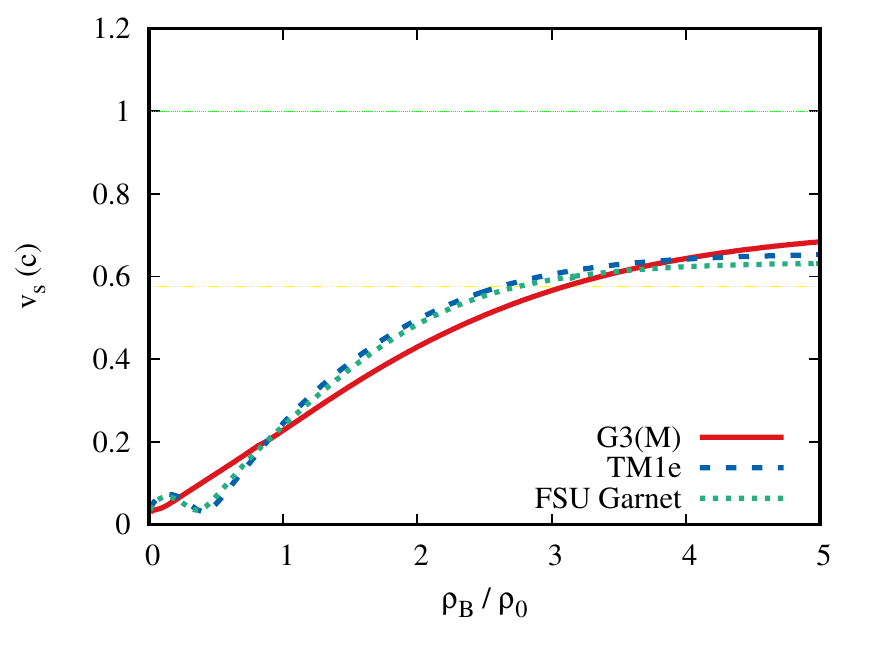}
  \caption{\label{fig3} (Color online) Sound velocity $v_s$ (c) (in units of the speed of light) as a function of $\rho_B /\rho_0$ calculated in the E--RMF model using the same parameter sets as used in Fig.~\ref{fig1}. The lower and upper bounds for the velocity of sound in dense matter are respectively represented by $v_s (c) =$ 1/$\sqrt{3}$ (yellow dashed line) and  $v_s(c) =$ 1 (green dashed line).}
\end{figure}
%
%

The difference between the G3(M) and both TM1e and FSU Garnet parameter sets is the G3(M) parameter set has the nonzero coupling constants of the $\eta_1$, $\eta_2$, $\eta_\rho$, $\alpha_1$, $\alpha_2$, $f_\omega$, $f_\rho$, $\beta_\sigma$, and $\beta_\omega$, whereas for the TM1e and FSU Garnet parameter sets, those coupling constants are set equal to zero as shown in Table~\ref{tab:model1}.

Besides the difference on those coupling constants, the difference is also given by the value of the $g_\delta$ coupling constant, which contribute to stiffen the EOS at high density as well as the symmetry energy at sub--saturation density. For the FSU Garnet and TM1e parameter sets, they have different value for the $m_\delta$.

The non--linear cross--coupling constants play important role for obtaining a better EOS for PNM that affect the particle fractions, neutrino cross section, and mean free path. Here I emphasize again that in the G3(M) parameter set, the non--linear cross--coupling constant of $\rho$ meson with $\omega$ meson $\eta_{2\rho}$ and the coupling constant of the $\rho$ meson $g_\rho$ are obtained by fine--tuning these parameters to fit the neutron--skin thickness of ${}^{208}$Pb from the PREX--2 data, making a crucial different between the G3(M) parameter set and the TM1e and FSU Garnet parameter sets.

Using the effective Lagrangian in Eq.~(\ref{eq:eqxenont1}), the energy density $\mathscr{E}$ and pressure $P$ of the dense matter can be determined using the standard procedure by solving the energy--momentum tensor,
\begin{eqnarray}
  T_{\mu \nu} = \sum_k \partial_\nu \phi_k \frac{\partial \mathscr{L}}{\partial \left( \partial^\mu \phi_k\right)} - g_{\mu \nu} \mathscr{L},
\end{eqnarray}
where $\phi_k$ are all the fields in the Lagrangian in Eq.~(\ref{eq:eqxenont1}). Thus, the energy density is obtained by taking the zeroth component of the energy--momentum tensor $\langle T_{00} \rangle$ that gives $\mathscr{E} = \langle T_{00} \rangle$, and the pressure is obtained from the energy--momentum tensor third component $\langle T_{kk} \rangle$ that gives $ P = \sum_{k}\frac{1}{3} \langle T_{kk} \rangle $. The energy density and pressure expressions for PNM are respectively given by
\begin{eqnarray}
  \label{eq:edenPNM}
  \mathscr{E} &=& \sum_{i=n,p} \frac{2}{(2\pi)^3} \int_0^{k_F^i} d^3 k \sqrt{\bm{k}_i^2 + M_i^{* 2}} + \rho_B g_\omega \omega - \frac{1}{24} \xi_0 g_\omega^2 \omega^4 \nonumber \\
  &+& \frac{1}{2} m_\sigma^2 \sigma^2 \left( 1 + \frac{\kappa_3}{3M_N} g_\sigma \sigma + \frac{\kappa_4}{12 M_N^2} g_\sigma^2 \sigma^2 \right) \nonumber \\
  &-& \frac{1}{2} m_\omega^2 \omega^2 \left( 1 + \frac{\eta_1}{M_N} g_\sigma \sigma + \frac{\eta_2}{2M_N^2} g_\sigma^2 \sigma^2 \right) \nonumber \\
  &-& \frac{1}{2} m_\rho^2 \rho^2 \left( 1 + \frac{\eta_\rho}{M_N} g_\sigma \sigma + \frac{\eta_{1 \rho}}{2M_N^2} g_\sigma^2 \sigma^2 \right)- \frac{\eta_{2 \rho}}{2} g_\rho^2 g_\omega^2 \rho^2 \omega^2 \nonumber \\
  &+& \frac{1}{2} \rho_3 g_\rho \rho + \frac{1}{2} m_\delta^2 \delta^2,
\end{eqnarray}
and
\begin{eqnarray}
  \label{eq:pressPNM}
  P &=& \sum_{i =n,p} \frac{1}{3} \frac{2}{(2\pi)^3} \int_0^{k_F^i} d^3 k \frac{\bm{k}_i^2}{\sqrt{\bm{k}_i^2 + M_i^2}} + \frac{1}{24} \xi_0 g_\omega^2 \omega^4 \nonumber\\
  &-& \frac{1}{2} m_\sigma^2 \sigma^2 \left( 1 + \frac{\kappa_3}{3M_N} g_\sigma \sigma + \frac{\kappa_4}{12 M_N^2}g_\sigma^2 \sigma^2 \right) \nonumber \\
  &+& \frac{1}{2} m_\omega^2 \omega^2 \left( 1 + \frac{\eta_1}{M_N} g_\sigma \sigma + \frac{\eta_2}{2M_N^2} g_\sigma^2 \sigma^2 \right)\nonumber \\
  &+& \frac{1}{2} m_\rho^2 \rho^2 \left( 1 + \frac{\eta_\rho}{M_N} g_\sigma \sigma + \frac{\eta_{1 \rho}}{2M_N^2} g_\sigma^2 \sigma^2 \right) +  \frac{\eta_{2 \rho}}{2} g_\rho^2 g_\omega^2 \rho^2 \omega^2 \nonumber \\
  &-& \frac{1}{2} m_\delta^2 \delta^2. 
\end{eqnarray}
With the Fermi momentum $k_F$ is defined by $\rho_B = \frac{k_F^3}{3\pi^2}$, and $\rho_3 = \rho_p - \rho_n$.

Using the energy density in Eq.~(\ref{eq:edenPNM}) and pressure in Eq.~(\ref{eq:pressPNM}), one can construct the relationship between $P$ and $\mathscr{E}$. Thus, the sound velocity $v_s(c)$ can be straightforwardly determined from the derivative of the $P$ with respect to the energy density $\mathscr{E}$, it then gives
\begin{eqnarray}
  \frac{v_s}{c} = \left( \frac{\partial P}{\partial \mathscr{E}} \right)^{\frac{1}{2}},
\end{eqnarray}
with $c$ is the speed of light.

Result for binding energy per nucleon $E_B/A$ calculated in the E--RMF model for the G3(M), TM1e and FSU Garnet parameter sets is shown in Fig.~\ref{fig1}. Figure~\ref{fig1} shows that the binding energy for the G3(M) parameter set is softer at low baryon density compared with the binding energy obtained for the TM1e and FSU Garnet parameter sets. However, at intermediate and high baryon densities, the binding energy for the G3(M) parameter set is rather stiffer than that obtained for the FSU Garnet parameter set and the same as that obtained for the TM1e parameter set. In addition, the $E_B/A$ for the G3(M) parameter set is in excellent agreement with the chiral effective field theory calculation with N$^3$LO 3N forces~\cite{Tews:2012fj}, in particular at higher densities.

Figure~\ref{fig2} shows the pressure for PNM with the G3(M), TM1e and FSU Garnet parameter sets compared with the experimental data~\cite{Danielewicz:2002pu}. The pressure for the G3(M) parameter set fits well with the asy--soft experimental data~\cite{Danielewicz:2002pu} at intermediate $\rho_B /\rho_0$. However, it overlaps with the asy--stiff experimental data~\cite{Danielewicz:2002pu} at higher $\rho_B/\rho_0$. At lower density ($\rho_B/ \rho_0 \lesssim 2$), the pressure for PNM with G3(M) parameter set has an excellent agreement with the chiral effective field theory calculation with N$^3$LO 3N forces~\cite{Tews:2012fj}.

The sound velocity calculated in the E--RMF model using the G3(M) parameter set is shown in Fig.~\ref{fig3}. Compared with the sound velocity for the TM1e and FSU Garnet parameter sets, the G3(M) parameter set has lower sound velocity at intermediate $\rho_B /\rho_0$, but higher sound velocity at very lower and higher $\rho_B / \rho_0$. All models predict the same sound velocity at around saturation density ($\rho_B / \rho_0 \simeq $ 1) and at $\rho_B / \rho_0 \simeq$ 3.7, which is a crossing point of the models. Also, all models satisfy the upper bound $v_s(c)$ constraint, meaning the models do not violate the causality.  

\begin{table}[t]
\caption{The complete G3(M)~\cite{PPBP21} that determined by readjusting to the PREX-2 data~\cite{Adhikhari21}, TM1e~\cite{Bao:2014lqa} and FSU Garnet~\cite{CP14} parameter sets. The nucleon mass $M_N$ is 939 MeV and all coupling constants are dimensionless. The unit of $k_3$ is in fm$^{-1}$.
}
\label{tab:model1}
\addtolength{\tabcolsep}{4.6pt}
\begin{tabular}{cccc} 
\hline \hline
Parameters & G3(M) & TM1e & FSU Garnet \\[0.2em] 
\hline
$m_s / M_N$   & 0.559 & 0.511 & 0.529 \\ 
$m_\omega /M_N$ & 0.832 & 0.783 & 0.833\\
$m_\rho /M_N$   & 0.820 & 0.770 & 0.812\\
$m_\delta /M_N$ & 1.043 & 0.980 & 0.000\\
$g_s /4\pi$    & 0.782 & 0.798 & 0.837\\
$g_\omega / 4 \pi$ & 0.923 & 1.004 & 1.091\\
$g_\rho / 4 \pi$ & 0.872 & 1.112 & 1.105 \\
$g_\delta / 4 \pi$ & 0.160 & 0.000 & 0.000\\
$k_3$ & 2.606 & -1.021 & 1.368 \\
$k_4$ & 1.694 & 0.124 & -1.397 \\
$\beta_\omega$ &-0.484 & 0.000 & 0.000\\
$\xi_0$ & 1.010 & 2.689 & 4.410\\
$\eta_1$ & 0.424 &0.000 & 0.000\\
$\eta_2$ & 0.114 &0.000 & 0.000\\
$\eta_\rho$ & 0.645 & 0.000 & 0.000\\
$\eta_{1 \rho}$ & 0.000 &0.000 & 0.000\\
$\eta_{2\rho}$ & 18.257 & 50.140 & 50.698\\
$\alpha_1$ & 2.000 & 0.000 & 0.000\\
$\alpha_2$ & -1.468 & 0.000 & 0.000\\
$f_\omega / 4$ & 0.220 &0.000 & 0.000\\
$f_\rho / 4$ & 1.239  &0.000 & 0.000\\
 $\beta_\sigma $ & -0.087 & 0.000& 0.000
\\ \hline \hline
\end{tabular}
\end{table}

\section{Neutrino scattering in dense matter}
\label{sec:interaction}

\subsection{Differential cross section}
%

Based on the weak interaction in the standard model (SM), Lagrangian density for the neutrino interaction with each constituent of matter is given by the current--current interaction and it has the form~\cite{SHM05b,Horowitz:1990it}
\begin{eqnarray}
  \label{eq:df1}
  \mathscr{L}_{int}^{j} &=& \tilde{G}_F \left[ \bar{\nu} \gamma^\mu (1-\gamma_5) \nu \right] \left( \bar{\psi} \Gamma_\mu^i \psi \right),
\end{eqnarray}
where $\tilde{G}_F = \frac{G_F}{\sqrt{2}}$, where $G_F = 1.023 \times 10^{-5} / M_N^2$ and $M_N$ is the nucleon mass. The nucleon vertex is given by $\Gamma_\mu^j = \gamma_\mu \left(C_V^j - C_A^j \gamma_5  \right)$ with $j = (n, p, e^{-}, \mu^{-})$ stands for the constituents of matter. For neutron, $C_V = -0.5$ and $C_A = - g_A/2$ and for proton, $C_V = 0.5 - 2 \sin^2 \theta_w$ and $C_A = g_A/2$, where $g_A = 1.260$ is the axial coupling constant and $\sin^2 \theta_w = 0.223$, respectively. For the electron, $C_V = 0.5 + 2 \sin^2 \theta_w$ and $C_A = 0.5$, whereas for the muon, $C_V = -0.5 + 2 \sin^2 \theta_w$ and $C_A = -0.5$. Further details of the values of $C_V$ and $C_A$ can be found in Ref.~\cite{Reddy:1997yr,SHM05b,HWSM04,Horowitz:1990it}.

For the charged--current absorption reactions, the interaction Lagrangian for the lepton and baryon in Eq.~(\ref{eq:df1}) can be rewritten as
\begin{eqnarray}
  \label{eq:df1a}
  \mathscr{L}_{int}^{(cc)} &=& \tilde{G}_F C \left[ \bar{\psi}_l \gamma^\mu (1-\gamma_5) \nu \right] \left( \bar{\psi} \Gamma_\mu^{(cc)} \psi \right),
\end{eqnarray}
where $\Gamma_\mu^{(cc)} = \gamma_\mu \left(g_V - g_A \gamma_5 \right)$ and $ \bar{\psi}_l$ are leptons. The $C$ is the Cabibbo factor with $C= \cos \theta_c$ for strangeness $\Delta S=0$ and $C= \sin \theta_c$ for $\Delta S = 1$. The values for $g_V$ and $g_A$ for the corresponding reactions can be found in Ref.~\cite{Reddy:1997yr,SHM05b,HWSM04,Horowitz:1990it}. Note that the DCRS for the neutral--current scattering has a similar expression as that for the charged--current absorption. The difference comes only from the values of the axial and vector coupling constants.

The neutrino differential cross section is straightforwardly derived from the Lagrangian in Eq.~(\ref{eq:df1}) and it gives~\cite{SHM05b,Horowitz:1990it}
\begin{eqnarray}
  \label{eq:df2}
  \frac{1}{V} \frac{d^3\sigma}{d^2 \Omega' dE_\nu'}  &=& - \frac{G_F}{32 \pi^2} \frac{E_\nu'}{ E_\nu} \Im \left[ L_{\mu \nu} \Pi^{\mu \nu} \right],
\end{eqnarray}
with $E_\nu$ and $E_\nu'$ are the initial and final neutrino energies, respectively. The neutrino tensor $L_{\mu \nu}$ can be defined by
\begin{eqnarray}
  L_{\mu \nu} &=& 8\left[ 2 k_\mu k_\nu + (k \cdot q) g_{\mu \nu} - (k_\mu q_\nu + q_\mu k_\nu) - i \epsilon_{\mu \nu \alpha \beta} k^{\alpha} q^{\beta} \right], \nonumber \\
\end{eqnarray}
where the four--momentum transfer is defined as $q = (q_0,\vec{q})$ and $k$ is the initial neutrino four--momentum. The polarization tensor $\Pi^{\mu \nu}$ for each target particles can be defined by~\cite{SHM05b,Horowitz:1990it}
\begin{eqnarray}
  \label{eq:df3}
  \Pi_{\mu \nu}^j &=& -i \int \frac{d^4p}{(2\pi)^4} Tr \left[ G^j(p) \Gamma_\mu^j G^j (p+q) \Gamma_\nu^j \right], 
\end{eqnarray}
where $p = (p_0,\bm{p})$ is the initial four--momentum of the target particles and $G^j (p)$ is the propagator of the target particle $j$, which is explicitly can be defined as
\begin{eqnarray}
  \label{eq:df4}
  G^{n,p} (p) &=& \left( p^*\!\!\!\!\!/ + M^* \right) \Bigg[ \frac{1}{[p^{*2} - M^{*2} + i \epsilon]} + \frac{i\pi}{E^*} \nonumber \\
    &\times& \delta \left( p_0^* - E^* \right) \Theta \left( p_F^{p,n} - |\bm{p}| \right) \Bigg], 
\end{eqnarray}
where the $E^* = \sqrt{\bm{p}^{*2} + M^{* 2}} = E + \Sigma_0$ is the effective nucleon energy and $M^* = M + \Sigma_s $ is the nucleon effective mass, where $\Sigma_s$ and $\Sigma_0$ are respectively the scalar and time--like self--energies. The nucleon effective momentum is defined as $\bm{p}^* = \bm{p} + \left( \frac{\bm{p}}{\mid \bm{p} \mid}\right) \Sigma_v$, where $|\bm{p}|$ is the three component--momentum of nucleon and $\Sigma_v$ is the spacelike self--energy. The $p_F^{p,n} = \sqrt{E_F^2 - M^{*2}}$ is the nucleon (proton and neutron) Fermi momentum.

The electron and muon propagators are taken the same as the free electron and muon propagators, respectively. The details of analytic derivations of the polarization tensors and the contractions of the leptonic and hadronic parts for the weak interaction as well as other quantities in Eq.~(\ref{eq:df2}) can be found in Refs.~\cite{Reddy:1997yr,Hutauruk:2010tn}.

\subsection{Neutrino mean free path}
%

In this section I present the NMFP of the neutrino scattering. The final expression for the inverse NMFP obtained by integrating the differential cross section of Eq.~(\ref{eq:df2}) over the energy transfer $q_0^{}$ and the three component--momentum transfer $\abs{\bm{q}}$ at a fixed baryon density can be obtained as~\cite{SHM05b,Horowitz:1990it}
\begin{align}
  \label{eq:df5}
  \lambda (E_\nu)^{-1} &= \int_{q_0^{}}^{2E_\nu - q_0^{}} d \abs{ \bm {q} } 
  \int_0^{2E_\nu} d q_0^{} \frac{\abs{\bm{q}}}{E'_\nu E_\nu} \frac{2\pi}{V} 
  \frac{d^3 \sigma}{d^2 \Omega' dE'_\nu},
\end{align}
where the final and initial neutrino energies are related as $E'_\nu = E_\nu - q_0^{}$. Further detailed explanations for the determination of the lower and upper limits of the integral can be found in Ref.~\cite{SHM05b,Horowitz:1990it}.

\section{Numerical results} \label{results}

\begin{figure*}[t]
  \centering\includegraphics[width=\textwidth]{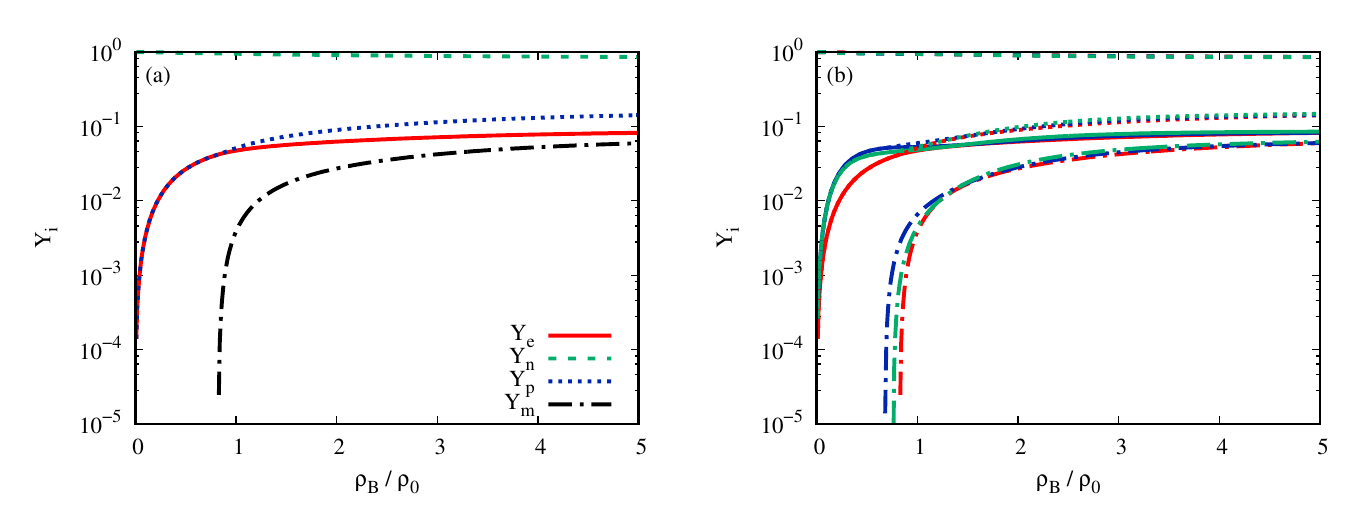}
  \caption{\label{fig4} (Color online) Particle fractions of the constituents of the $\beta-$stable nuclear matter as a function of $\rho_B /\rho_0$ calculated in the E--RMF model for (a) the G3(M) parameters set and (b) for the G3(M), TM1e and FSU Garnet parameter sets which are represented by different colors. The $Y_e$, $Y_n$, $Y_p$, and $Y_m$ represent respectively the electron, neutron, proton and muon fractions.}
\end{figure*}
%
%
Here the numerical results for the particle fractions of the constituents of $\beta$--stable matter, DCRS of neutrino and NMFP for the G3(M) parameter set are presented. The neutrino DCRS and NMFP are calculated with fixed values of the three component--transferred momentum $\abs{\bm{q}} = 2.5$ MeV and initial neutrino energy $E_\nu = 5$ MeV.

Result for the particle fractions of electrons, neutrons, protons, and muons as a function of $\rho_B / \rho_0$ for only the G3(M) parameter set is shown in Fig.~\ref{fig4}(a). The particle fractions for all parameter sets are shown in Fig.~\ref{fig4}(b). The particle fractions of neutrons, protons, and electrons for the G3(M) parameter set are almost unchanged compared with that obtained for the TM1e and FSU Garnet parameter sets. However, the appearing of muons for the G3(M) parameter set is rather longer than that obtained for the TM1e and FSU Garnet parameter sets, as shown in Fig.~\ref{fig4}(b).

\begin{figure*}[t]
  \centering\includegraphics[width=\textwidth]{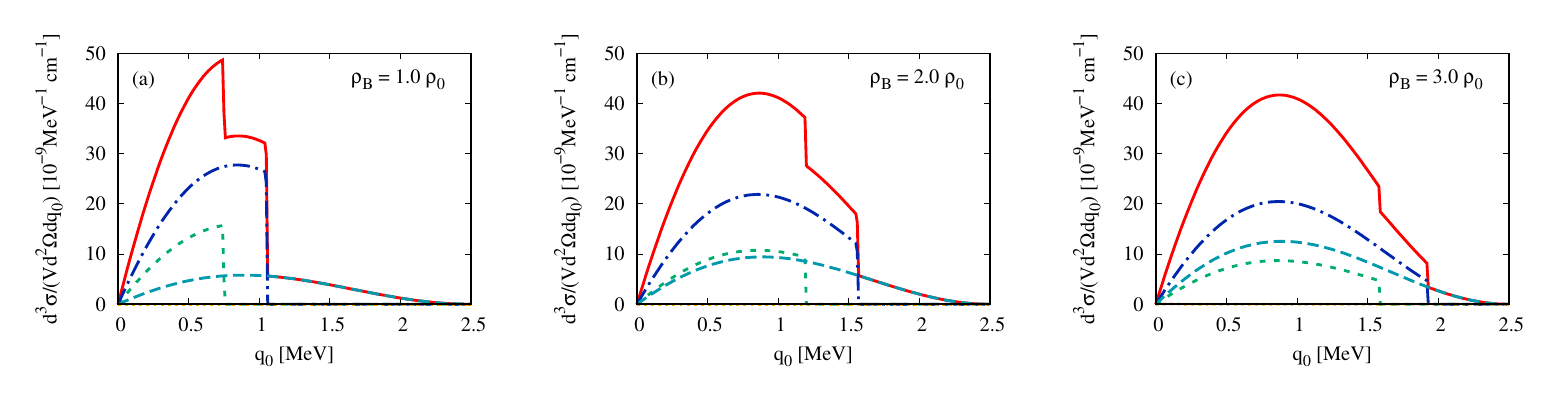}
  \caption{\label{fig5} (Color online) DCRS of neutrino as a function of $q_0$ for (a) $\rho_B = 1.0 ~\rho_0$ (b) $\rho_B = 2.0~\rho_0$ and (c) $\rho_B = 3.0~\rho_0$ for the G3(M) parameter set at $\abs{\bm{q}} = 2.5$ MeV and $E_\nu = 5 $ MeV. The solid, dotted-slashed, dashed, long-dashed and dotted lines are the differential cross section of neutrino for total (electrons + neutrons + protons + muons), neutrons, protons, electrons and muon, respectively.}
\end{figure*}
%
%

\begin{figure*}[t]
  \centering\includegraphics[width=\textwidth]{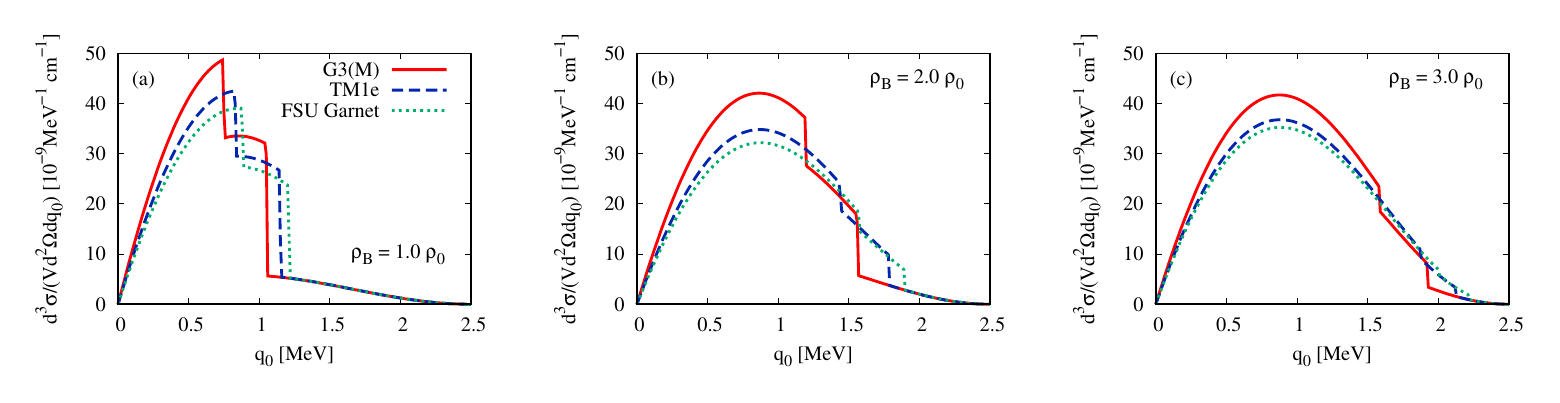}
  \caption{\label{fig6} (Color online) Total DCRS of neutrino as a function of $q_0$ for the same parameters set as in Fig.~\ref{fig1} for (a) $ \rho_B = 1.0~\rho_0$ (b) $\rho_B = 2.0~\rho_0$ and (c) $\rho_B = 3.0~\rho_0$.}
\end{figure*}
%
%

\begin{figure*}[t]
  \centering\includegraphics[width=\textwidth]{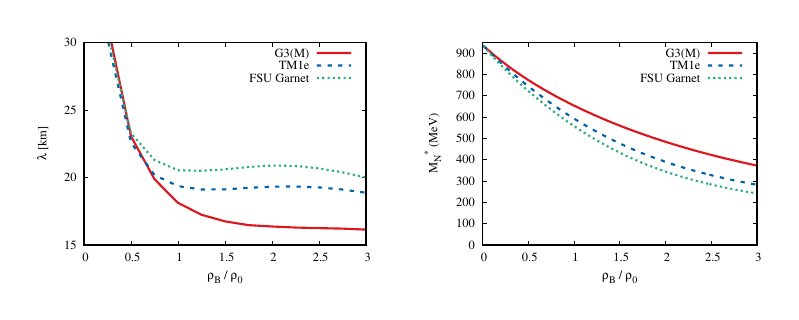}
  \caption{\label{fig7} (Color online) NMFP (left panel) and nucleon effective mass $M_N^*$ (right panel) as a function of $\rho_B /\rho_0$ for the G3(M), TM1e and FSU Garnet parameter sets.}
\end{figure*}
%

Next, the DCRS of neutrino for the G3(M) parameter set as a function of energy transferred momentum $q_0$ for different baryon densities (a) $\rho_B = 1.0~\rho_0$ (b) $\rho_B = 2.0~\rho_0$ and (c) $\rho_B = 3.0~\rho_0$ as in Fig.~\ref{fig5}. The patterns of the DCRS of neutrino for each constituent significantly change as the baryon density increases. Consequently, it leads to the change of the shape and magnitude of the total DCRS.

Compared with total DCRS of neutrino for the TM1e and FSU Garnet parameter sets, the G3(M) parameter set has higher value of the cross section as shown in Figs.~\ref{fig6}(a)--\ref{fig6}(c). It shows that the DCRS of neutrino is sensitive to the parameter set used. However, in general, the patterns of the DRCS of neutrino for different parameters sets are rather the same.

The change of DCRS of neutrino for each parameter set affects the NMFP as shown in Fig.~\ref{fig7}. The left panel of Fig.~\ref{fig7} shows the NMFP for the G3(M) parameter set is lower than that obtained for the TM1e and FSU Garnet parameter sets. However, the NMFP for each parameters set decreases as the $\rho_B /\rho_0$ increases. Note that the higher NMFP is given by the FSU Garnet parameter set. Increasing DCRS or decreasing NMFP is expected due to the nucleon effective mass $M_N^*$ as shown in the right panel of Fig.~\ref{fig7}. The $M_N^*$ for the G3(M) parameter set is higher than that obtained for the FSU Garnet and TM1e parameter sets in particular at higher densities.

\section{Summary} \label{summary}
%

To summarize, I have studied the implications of the G3(M) parameter set that constrained by PREX--2 data on the equation of state, the particle fractions of the constituents of the matter, differential cross section of neutrino, and NMFP in the E--RMF model.

One finds that the binding energy per nucleon for the G3(M) parameter set is softer at low baryon density compared with that obtained for the TM1e and FSU Garnet parameter sets. In contrast, at higher baryon density, the binding energy for the G3(M) parameter set is stiffer than that obtained for the FSU Garnet parameter set and the same as that obtained for the TM1e parameter set. The $E_B/A$ for the G3(M) parameter set fits well with the result of the chiral effective field theory calculation with N$^3$LO 3N forces~\cite{Tews:2012fj}, in particular at higher densities.

Result for the pressure for pure neutron matter, one finds that the pressure for pure neutron matter for the G3(M) parameter set fits well with the asy--soft experimental data at intermediate baryon density. However, at higher baryon density, it has a good agreement with the asy--stiff data.

Result for the sound velocity, one finds that the sound velocity for the G3(M) parameter set is lower than that obtained for the TM1e and FSU Garnet parameter sets at intermediate $\rho_B / \rho_0$ but it is higher at higher $\rho_B \ /\rho_0$.

One finds the differential cross section of neutrino for different densities with the G3(M) parameter set has higher value compared with that obtained for the TM1e and FSU Garnet parameter sets and it affects to decrease the neutrino mean free path.

\begin{acknowledgments}
P.T.P.H. thanks A. Sulaksono for valuable conversation and discussion and Seung--Il Nam for useful conversation. This work was supported by the National Research Foundation of Korea (NRF) grant funded by the Korea government (MSIT) No.~2018R1A5A1025563 and No.~2019R1A2C1005697.
\end{acknowledgments}


\begin{thebibliography}{10}

\bibitem{Yakovlev:2000jp}
D.~G.~Yakovlev, A.~D.~Kaminker, O.~Y.~Gnedin and P.~Haensel,
\newblock Neutrino emission from neutron stars,
\newblock Phys. Rept. \textbf{354}, 1 (2001).

\bibitem{Burrows:1987zz}
A.~Burrows and J.~M.~Lattimer,
\newblock Neutrinos from SN 1987A,
\newblock Astrophys. J. Lett. \textbf{318}, L63-L68 (1987).

\bibitem{Hutauruk:2020mhl}
P.~T.~P.~Hutauruk, A.~Sulaksono and K.~Tsushima,
\newblock Effect of neutrino magnetic moment and charge radius on the neutrino mean free path in dense matter with medium modifications of the nucleon form factors,
\newblock [arXiv:2009.08781 [hep-ph]].

\bibitem{Hutauruk:2019ptu}
P.~T.~P.~Hutauruk, Y.~Oh and K.~Tsushima,
\newblock Effects of medium modifications of nucleon form factors on neutrino scattering in dense matter,
\newblock JPS Conf. Proc. \textbf{26}, 024031 (2019).

\bibitem{Hutauruk:2018cgu}
P.~T.~P.~Hutauruk, Y.~Oh and K.~Tsushima,
\newblock Impact of medium modifications of the nucleon weak and electromagnetic form factors on the neutrino mean free path in dense matter,
\newblock Phys. Rev. D \textbf{98}, no.1, 013009 (2018).

\bibitem{Reddy:1997yr}
S.~Reddy, M.~Prakash and J.~M.~Lattimer,
\newblock Neutrino interactions in hot and dense matter,
\newblock Phys. Rev. D \textbf{58}, 013009 (1998).

\bibitem{Niembro:2001hd}
R.~Niembro, P.~Bernardos, M.~Lopez-Quelle and S.~Marcos,
\newblock Neutrino cross-section and mean free path in neutron stars in the framework of the Dirac-Hartree-Fock approximation,
\newblock Phys. Rev. C \textbf{64}, 055802 (2001).

\bibitem{Adhikhari21}
D.~Adhikari \textit{et al.} [PREX],
\newblock Accurate Determination of the Neutron Skin Thickness of $^{208}$Pb through Parity-Violation in Electron Scattering,
\newblock Phys. Rev. Lett. \textbf{126}, no.17, 172502 (2021).

\bibitem{Abrahamyan2012}
S.~Abrahamyan, Z.~Ahmed, H.~Albataineh, K.~Aniol, D.~S.~Armstrong, W.~Armstrong, T.~Averett, B.~Babineau, A.~Barbieri and V.~Bellini, \textit{et al.}
\newblock Measurement of the Neutron Radius of 208Pb Through Parity-Violation in Electron Scattering,
\newblock Phys. Rev. Lett. \textbf{108}, 112502 (2012).
  
\bibitem{PPBP21}
J.~A.~Pattnaik, R.~N.~Panda, M.~Bhuyan and S.~K.~Patra,
\newblock Constraining the relativistic mean-field models from PREX-2 data: Effective forces revisited,
\newblock  [arXiv:2105.14479 [nucl-th]].

\bibitem{RFHP21}
B.~T.~Reed, F.~J.~Fattoyev, C.~J.~Horowitz and J.~Piekarewicz,
\newblock Implications of PREX-2 on the Equation of State of Neutron-Rich Matter,
\newblock Phys. Rev. Lett. \textbf{126}, no.17, 172503 (2021).

\bibitem{Piekarewicz21}
J.~Piekarewicz,
\newblock Implications of PREX-2 on the electric dipole polarizability of neutron rich nuclei,
\newblock [arXiv:2105.13452 [nucl-th]].

\bibitem{Kumara:2017bti}
B.~Kumar, S.~K.~Singh, B.~K.~Agrawal and S.~K.~Patra,
\newblock New parameterization of the effective field theory motivated relativistic mean field model,
\newblock Nucl. Phys. A \textbf{966}, 197-207 (2017).

\bibitem{Agrawal:2012rx}
B.~K.~Agrawal, A.~Sulaksono and P.~G.~Reinhard,
\newblock Optimization of relativistic mean field model for finite nuclei to neutron star matter,
\newblock Nucl. Phys. A \textbf{882}, 1-20 (2012).

\bibitem{Furnstahl:1995zb}
R.~J.~Furnstahl, B.~D.~Serot and H.~B.~Tang,
\newblock Analysis of chiral mean field models for nuclei,
\newblock Nucl. Phys. A \textbf{598}, 539-582 (1996).

\bibitem{Bao:2014lqa}
S.~S.~Bao, J.~N.~Hu, Z.~W.~Zhang and H.~Shen,
\newblock Effects of the symmetry energy on properties of neutron star crusts near the neutron drip density,
\newblock Phys. Rev. C \textbf{90}, no.4, 045802 (2014).

\bibitem{CP14}
W.~C.~Chen and J.~Piekarewicz,
\newblock Searching for isovector signatures in the neutron-rich oxygen and calcium isotopes,
\newblock Phys. Lett. B \textbf{748}, 284-288 (2015).

\bibitem{Tews:2012fj}
I.~Tews, T.~Kr\"uger, K.~Hebeler and A.~Schwenk,
\newblock Neutron matter at next-to-next-to-next-to-leading order in chiral effective field theory,
\newblock Phys. Rev. Lett. \textbf{110}, no.3, 032504 (2013).

\bibitem{Danielewicz:2002pu}
P.~Danielewicz, R.~Lacey and W.~G.~Lynch,
\newblock Determination of the equation of state of dense matter,
\newblock Science \textbf{298}, 1592-1596 (2002).

\bibitem{Furnstahl:1996wv}
R.~J.~Furnstahl, B.~D.~Serot and H.~B.~Tang,
\newblock A Chiral effective Lagrangian for nuclei,
\newblock Nucl. Phys. A \textbf{615}, 441-482 (1997)
[erratum: Nucl. Phys. A \textbf{640}, 505-505 (1998)].

\bibitem{SHM05b}
A.~Sulaksono, P.~T.~P. Hutauruk, and T.~Mart,
\newblock Isovector-channel role of relativistic mean field models in the neutrino mean free path,
\newblock Phys. Rev. C \textbf{72}, 065801 (2005).
  
\bibitem{HWSM04}
P.~T.~P. Hutauruk, C.~K. Williams, A.~Sulaksono, and T.~Mart,
\newblock Neutron fraction and neutrino mean free path predictions in relativistic mean field models,
\newblock Phys. Rev. C \textbf{70}, 068801 (2004).

\bibitem{Horowitz:1990it}
C.~J.~Horowitz and K.~Wehrberger,
\newblock Neutrino neutral current interactions in nuclear matter,
\newblock Nucl. Phys. A \textbf{531}, 665-684 (1991).

\bibitem{Hutauruk:2010tn}
P.~T.~Hutauruk,
\newblock Neutrino Mean Free Path in Neutron Star,
\newblock [arXiv:1007.4007 [nucl-th]].
 
\end{thebibliography}
\end{document}